\documentclass[prl, superscriptaddress, twocolumn, showpacs]{revtex4}  
\usepackage{amssymb}
\usepackage{amsmath}
\usepackage{epsfig}
\usepackage{epstopdf}
\usepackage{bbm}
\usepackage{graphicx}
\usepackage{color}
\begin{document}

\author{Philipp Werner}
\affiliation{Theoretische Physik, ETH Zurich, 8093 Z{\"u}rich, Switzerland}
\author{Andrew J. Millis}
\affiliation{Columbia University, 538 West, 120th Street, New York, NY 10027, USA}

\title{Dynamical Screening in Correlated Electron Materials} 
\date{January 8, 2010}

\hyphenation{}

\begin{abstract}
We present an efficient method for incorporating the dynamical effects of the screening of the Hubbard $U$ by electronic degrees of freedom in the solid into the single site dynamical mean field approximation. The formalism is illustrated by model system calculations which capture the essential features of the frequency dependent interactions proposed for Gd, Ni, SrVO$_3$ and other compounds. Screening leads to shifts in the metal-insulator phase boundary, changes in the spectral function near  the  Mott-Hubbard gap edge and to a renormalization of the quasiparticle weight. Hubbard bands are generically neither separated by the screened nor the unscreened interaction energy, implying that the common practice of extracting the Hubbard $U$ from the energies of features in photoemission and inverse photoemission spectra requires reexamination. 
\end{abstract}

%\pacs{ 71.10.Fd, 02.70.Ss, 71.27.+a, 71.30.+h}

\maketitle

`Strongly correlated electron systems' are a central topic in electronic condensed matter physics \cite{Imada98}. The low energy physics of these systems is typically described by an effective Hamiltonian which models the behavior of relatively localized $d$ or $f$ orbitals and is obtained (at least notionally)  via a ``downfolding" procedure in which other degrees of freedom are integrated out.  A crucial aspect of the effective Hamiltonian is an interaction which acts to suppress local number fluctuations. This interaction is typically parametrized by a number, the ``Hubbard $U$''.  However, $U$ is generically dynamical:  a density fluctuation in a correlated orbital produces electric fields, which other degrees of freedom will act to screen, resulting in a frequency-dependent renormalization. Screening has been observed as  a shift in excitation energies in experiments comparing the surface of solid  C$_{60}$  to C$_{60}$ films on silver  \cite{Hesper97} and  has been computed using variations of the `random phase approximation' (RPA) \cite{Aryasetiawan04, Aryasetiawan06, Aryasetiawan08}. The renormalizations are found to be strong in many cases. In Gd, $U(\omega)$ rises from a static value of about 6.5~eV to about 17~eV as the frequency $\omega $ is raised from $0$ to $\sim 3$~eV, while in Ce, $U(\omega)$ changes from $\sim 3.5$~eV to $\sim 7$~eV as $\omega$ is increased from $0$ to $\sim 4$~eV \cite{Aryasetiawan06}.  

While the study of the strong correlation effects induced by an instantaneous interaction is well advanced thanks to  the development of dynamical mean field theory (DMFT) \cite{Georges96}, our ability to treat frequency dependent interactions has been limited.  The most widely used method for solving the DMFT equations has been the Hirsch-Fye algorithm \cite{Hirsch86}, which is based on a time-discretization and decoupling of interaction terms by auxiliary fields. Frequency dependent interactions lead to a proliferation of decoupling fields which become prohibitively expensive to sample. A treatment of  screening effects within exact diagonalization, numerical renormalization group and other  Hamiltonian based methods requires the explicit introduction of many bosonic modes, leading to a Hilbert space which is too large to be handled numerically. Considerations of this sort have led to the belief (see e.g.~Ref.~\cite{Aryasetiawan04}) that DMFT simulations with frequency dependent interactions are exceedingly difficult. 

Here we show that this is not the case: the recently developed `hybridization expansion'  diagrammatic quantum Monte Carlo method \cite{Werner06,Werner06Kondo} can be used to treat models with an arbitrary frequency dependence of  the on-site repulsion $U(\omega)$ at negligible additional computational cost,  opening  the door to a systematic investigation of screening effects in correlated electron materials. We begin our discussion by recalling that  the downfolded models used to describe the correlated degrees of freedom in a transition metal or actinide involve a one electron part   and an interaction part. The parameters describing the one-electron physics are obtained by projecting a band theory calculation onto a set of distinguished (``$d$")   orbitals and are in principle energy dependent. If the $d$ orbitals are correctly chosen the energy dependence is negligible \cite{Zurek05,Kotliar06} so  the one electron part may be modelled as a tight-binding-like Hamiltonian $H_\text{band}$. 

The interaction part is obtained (see, e.g. Ref.~\onlinecite{Aryasetiawan04}) by  screening  the   bare Coulomb interaction $e^2/|r-r{'}|$ with  real and virtual transitions involving the orbitals which are integrated out,  projecting the result onto the $d$ manifold and retaining only the fully site-diagonal terms. One finds two kinds of terms: an instantaneous interaction  and a screening contribution. The instantaneous interaction may be represented as a Hamiltonian term $H_\text{int}$  which  takes the usual Slater-Kanamori form $H_\text{int}=\sum_i\frac{1}{2}U{\hat N}_i^2+....$ with ${\hat N}_i$ the number operator for electrons in the $d$ manifold on site $i$ and the ellipsis denoting exchange, `pair hopping' and other terms which involve operators such as the spin and angular momentum which   commute with ${\hat N}$. The screening  contribution couples the site densities at different times and is expressed as a contribution $S_W=\frac{1}{2}\int d\tau d\tau{'}N(\tau) W(\tau-\tau{'})N(\tau{'})$ to the effective action. The screening function $W(\tau)$  is negative, depends only on one time (or frequency) argument and couples only to the site density $N_i$,  reflecting the physics of dynamical screening.  It is generated by a screening spectral function $\text{Im}W(\omega)$:
\begin{eqnarray}
W(\tau)&=&\int_0^\infty \frac{d\omega_0}{\pi}\text{Im} W(\omega_0)W_{\omega_0}(\tau),\label{Wtau}
\label{Womegadef}
\end{eqnarray}
with $W_{\omega_0}(\tau)=\cosh\big[\big(\tau-\frac{\beta}{2}\big)\omega_0\big]/\sinh\big[\frac{\omega_0\beta}{2}\big]$  for $0\leq \tau \leq \beta$ and $W_{\omega_0}(\tau)=W_{\omega_0}(\tau+n\beta)$ ($n$ integer). Other interaction parameters  acquire only a very weak frequency dependence \cite{Aryasetiawan08} because they involve operators corresponding  to higher multipoles which are weakly screened.

\begin{figure}[t]
\begin{center}
\includegraphics[angle=0, width=0.8\columnwidth]{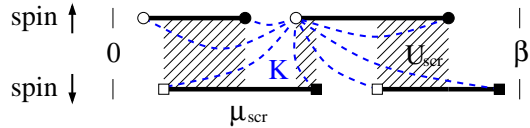}
\caption{
Illustration of an order $n=4$ diagram for the one orbital Hubbard model. Empty (full) circles and squares represent $V^\dagger$ ($V$) hybridization events. Dashed lines indicate interactions $K(\tau)$ connecting all pairs of hybridization events.}   
\label{diagram}
\end{center}
\end{figure}

We recast the time dependent interaction in Hamiltonian form by using  Hubbard-Stratonovich techniques to introduce boson operators whose spectrum reproduces $\text{Im} W$. We define $\lambda_{\omega_0}^2=-\pi \text{Im} W(\omega_0)$ and at each site $i$ and each frequency $\omega_0$ we employ the identity (``*" denotes integration over time) $e^{\frac{1}{2}\lambda_{\omega_0} N*W_{\omega_0}*N\lambda_{\omega_0}}= \int {\cal D}\phi_{\omega_0} e^{-\phi_{\omega_0} * W_{\omega_0}^{-1}*\phi_{\omega_0}-\sqrt{2}\phi_{\omega_0}*N\lambda_{\omega_0}}$. Noting that  $W^{-1}_{\omega_0}=(-\partial_\tau^2+\omega_0^2)/(2\omega_0)$ (the periodicity under $\tau\rightarrow \tau+n\beta$ implies a derivative discontinuity at $\tau=n\beta$) and identifying $\partial_\tau \phi_{\omega_0}$ as $\omega_0$ times the momentum $\Pi_{\omega_0}$ conjugate to $\phi_{\omega_0}$ we see that the time dependent interaction is equivalent to a Hamiltonian model in which the site density is coupled via  $\lambda_{\omega_0}$ to a set of oscillators,
\begin{equation}
H_\text{screen}=\sum_i\int_0^\infty \!\!\!d\omega_0 \left[\frac{\omega_0}{2}\Big(\hat\Pi_{\omega_0,i}^2+\hat \phi_{\omega_0,i}^2\Big) +\sqrt{2}\lambda_{\omega_0} \hat \phi_{\omega_0,i} {\hat N}_i\right].
\label{Hscreen}
\end{equation}
To solve the Hamiltonian $H=H_\text{band}+H_\text{int}+H_\text{screen}$ we employ dynamical mean field theory \cite{Georges96,Kotliar06} which computes the solution to the correlated electron problem from the solution of an auxiliary quantum impurity model with interactions given by the local interactions of the original model and an impurity-bath hybridization term $H_\text{hyb}=c^\dagger V d+\text{H.c.}$ with $V$ determined by a self-consistency condition. Following Ref.~\cite{Werner06Kondo} we solve the impurity model by expanding  the partition function in powers $V$, collecting diagrams into  determinants of  matrices of hybridization functions and using a Monte Carlo procedure to sample the resulting sum of determinants.  As in Ref.~\cite{Werner07Holstein} we treat the electron-boson coupling by a canonical transformation which shifts $\hat \phi_{\omega_0}$ by $\sqrt{2}\lambda_{\omega_0}{\hat N}/\omega_0$,  changes the instantaneous interaction $U$ to the screened value  
\begin{equation}
U_\text{scr}=U+2\int_0^\infty \frac{d\omega_0}{\pi} \frac{\text{Im}W(\omega_0)}{\omega_0}<U,
\label{Umod}
\end{equation}  
and shifts the chemical potential $\mu$ to $\mu_\text{scr}=\mu+\int_0^\infty \frac{d\omega_0}{\pi} \frac{\text{Im}W(\omega_0)}{\omega_0}$.  The transformation also multiplies $d$, $d^\dagger$ by $e^{is\int_0^\infty d\omega_0 \Pi_{\omega_0}\sqrt{2}\lambda_{\omega_0}/\omega_0}$ ($s=1$ for $d^\dagger$ and $s=-1$ for $d$). The result is that  a term in the hybridization expansion with $2n$ hybridization events at times $0\le \tau_1<\tau_2<\ldots<\tau_{2n}<\beta$ acquires an extra weight factor  $w_\text{screen}(\{\tau_i\})$ given by an exponential of correlators of noninteracting boson operators, which evaluates to: 
\begin{align} 
&w_\text{screen}(\{\tau_i\})=e^{\sum_{2n\geq i>j\geq1}s_is_jK(\tau_i-\tau_j)},\label{wscreen}\\
K(\tau)&=\int_0^\infty \frac{d\omega_0}{\pi}\frac{\text{Im}W(\omega_0)}{\omega_0^2}[W_{\omega_0}(\tau)-W_{\omega_0}(0)].
\label{Ktau}
\end{align}
It follows from Eqs.~(\ref{wscreen}) and (\ref{Ktau}) and the illustration in Fig.~\ref{diagram} that the screening contribution amounts to a non-local interaction between all pairs of hybridization events.  
Since the computational bottleneck is the handling of the hybridization matrix determinants, the additional weight factor $w_\text{screen}$ does not significantly slow down the simulation, which remains very efficient \cite{Gull07}. 

We now illustrate the method with single-site, single-orbital  DMFT calculations for a semi-circular density of states of bandwidth $D=4$ and  inverse temperature $\beta=50$.  All calculations are performed in the paramagnetic phase and for half filling. We consider two screening functions: (i) $K(\tau)=-(\lambda/\omega_0)^2(\cosh((\beta/2-\tau)\omega_0)-\cosh(\beta \omega_0/2))/\sinh(\beta\omega_0/2)$ corresponding to a delta-function $\text{Im}W(\omega)=-\lambda^2\pi\left(\delta(\omega-\omega_0)-\delta(\omega+\omega_0)\right)$ and (ii) $K(\tau)=\alpha\ln[1+\beta\omega_c\sin(\pi\tau/\beta)/\pi)]$ corresponding to ``Ohmic" screening $\text{Im} W(\omega)\sim -\alpha\pi \omega$ at small $\omega$, with an ultraviolet cutoff at $\omega_c$. Model (i) implies that the screened interaction is  $ U_\text{scr} = U-2\lambda^2/\omega_0$ and $\text{Re}W(\omega)=2\lambda^2\omega_0/(\omega^2-\omega_0^2)$; for model (ii) $ U_\text{scr}=U-2\alpha\omega_c$ and $\text{Re}W(\omega)=\alpha\omega\ln|(\omega_c+\omega)/(\omega_c-\omega)|-2\alpha\omega_c$. In principle the coupling strengths can be made large enough to drive $U_\text{scr}$ negative (overscreening), but in the absence of phonons overscreening is not believed to occur in real materials, limiting the range of physically relevant  couplings to $\lambda<\sqrt{\omega_0U/2}$ or $\alpha<U/(2\omega_c)$. Model (i) is a rough represention of results obtained from constrained RPA calculations for Gd and SrVO$_3$ and model (ii) captures a characteristic feature of paramagentic Ni \cite{Aryasetiawan04, Aryasetiawan06, Aryasetiawan08}. 

\begin{figure}[t]
\begin{center}
\includegraphics[angle=-90, width=0.575\columnwidth]{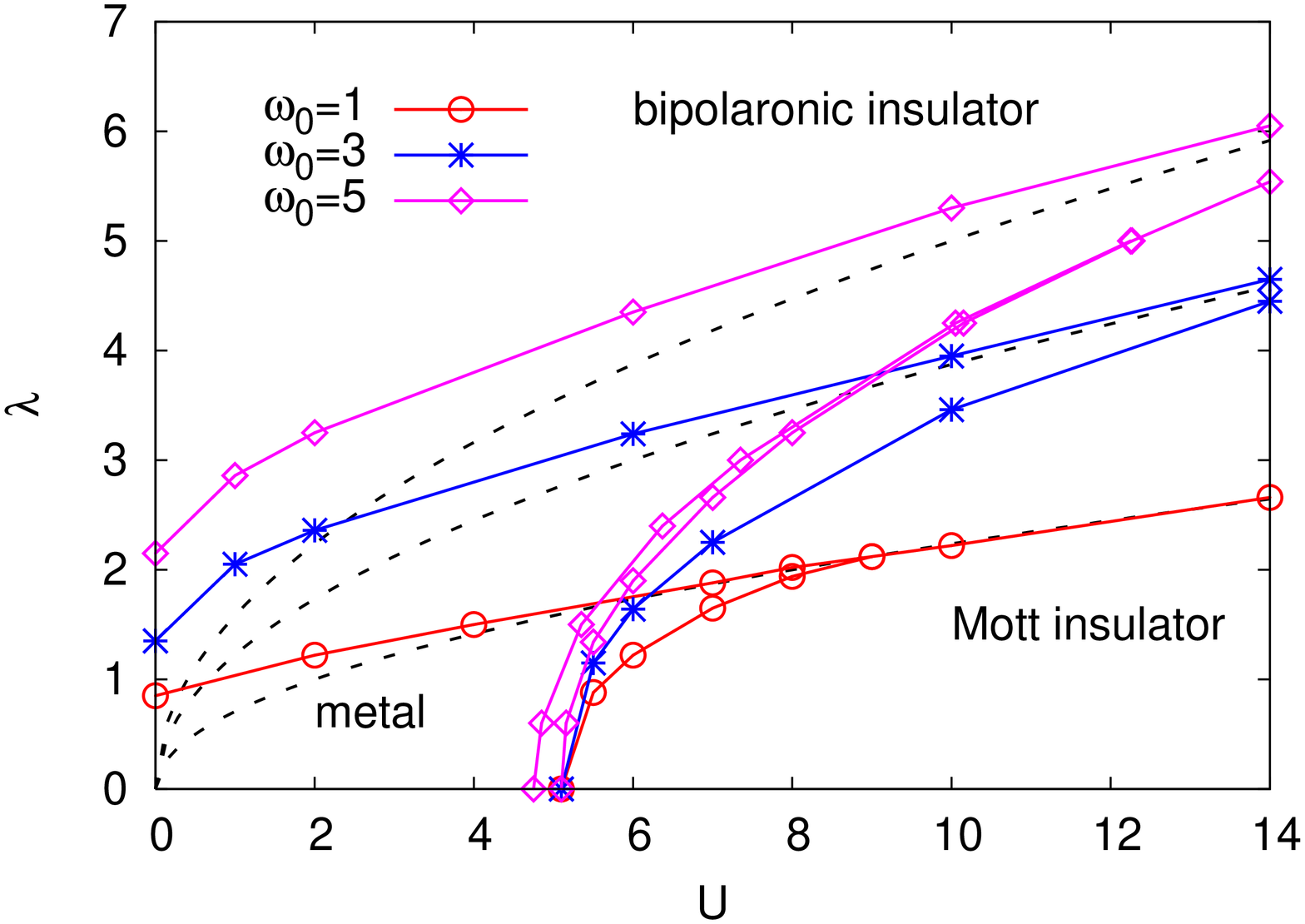}
\includegraphics[angle=-90, width=0.41\columnwidth]{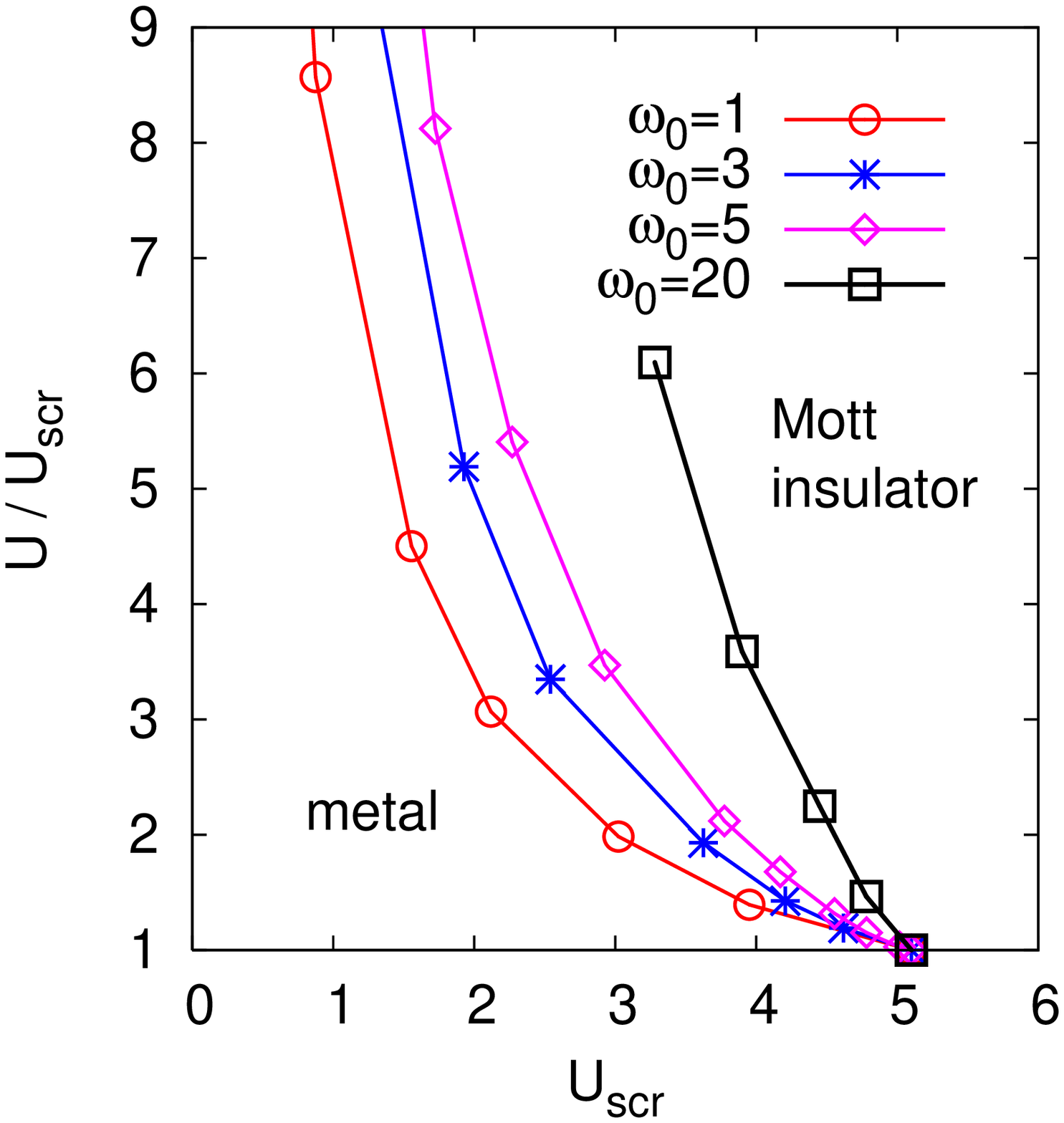}
\includegraphics[angle=-90, width=0.575\columnwidth]{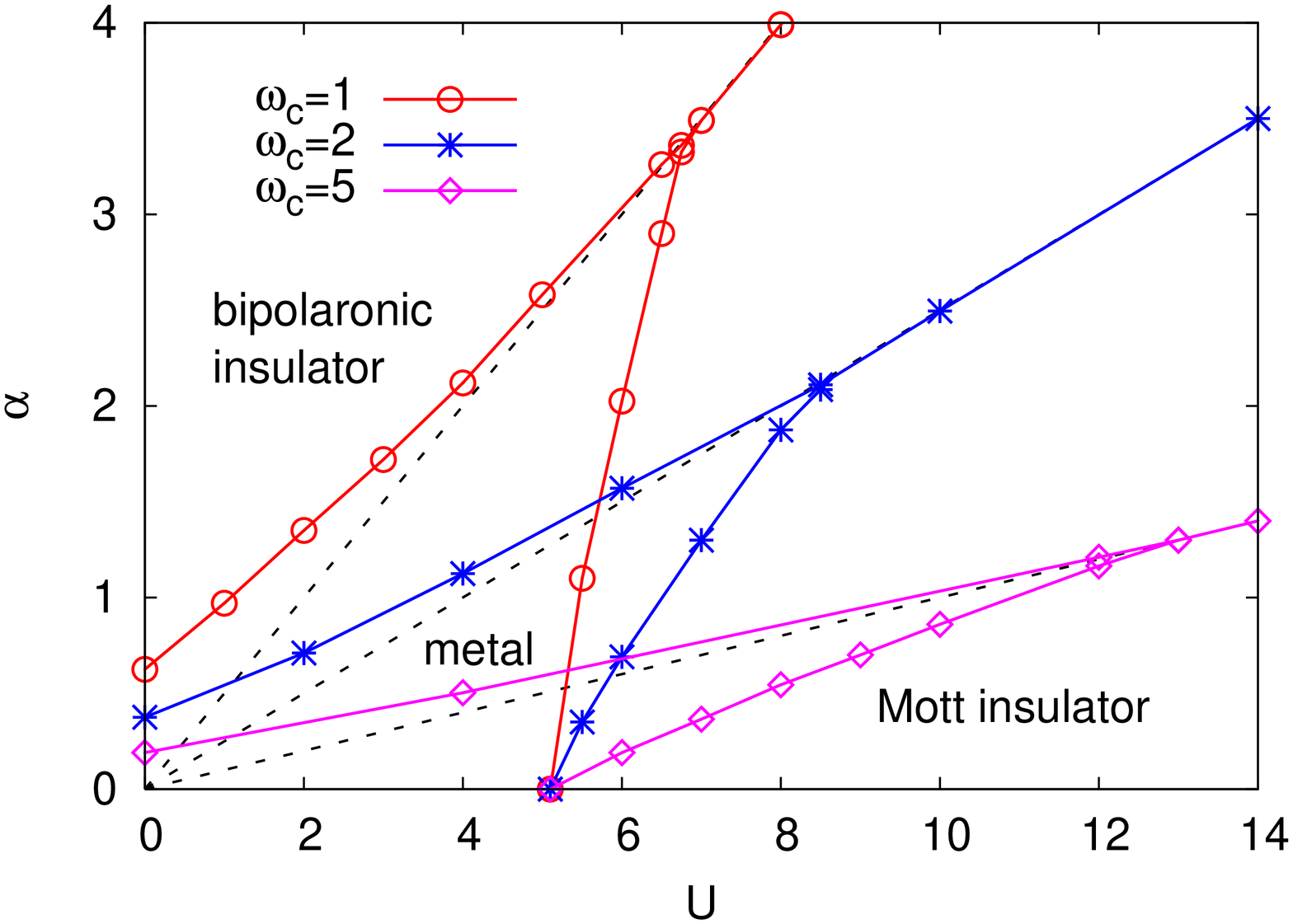}
\includegraphics[angle=-90, width=0.41\columnwidth]{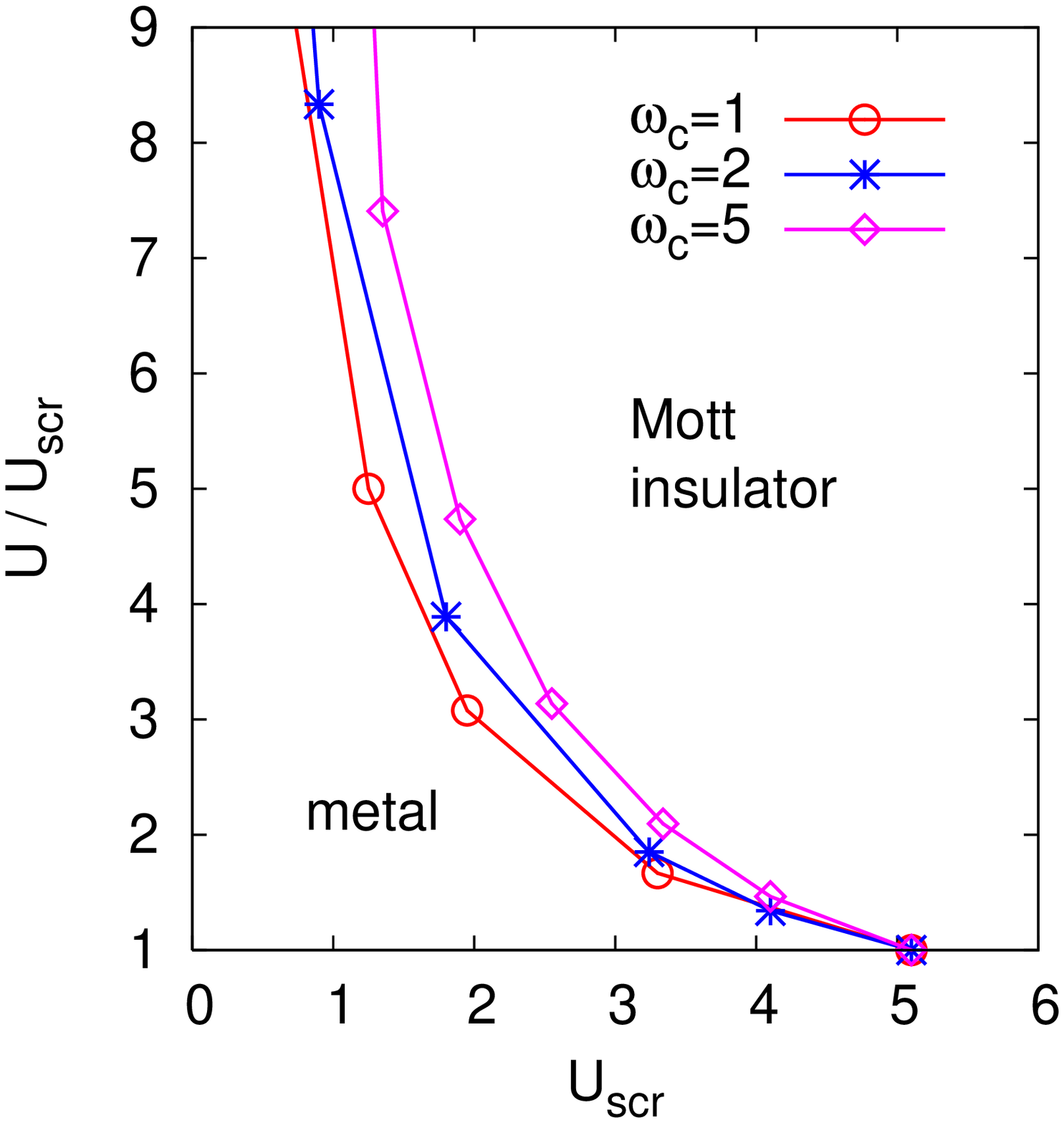}
\caption{Left panels: metal-insulator phase diagram in the space of bare interaction $U$ and screening coupling strength $\lambda$ ($\alpha$) for indicated values of the screening frequency $\omega_0$ ($\omega_c$). Upper panel: model (i); lower panel, model (ii).  Dashed lines: screening strength at which the screened interaction $U_\text{scr}$ 
changes sign. Right panels: phase diagram in the space of screened and bare interaction. }   
\label{phasediagram_i}
\end{center}
\end{figure}

The calculated metal-insulator phase diagrams for our two models are  shown in the  left-hand panels of Fig ~\ref{phasediagram_i}. At our simulation temperature the critical $U$ for the metal insulator transition of the unscreened model is $U_{c2}(\beta=50)\approx 5.1$ and the transition is  first-order \cite{Georges96}. As expected on physical grounds, increasing the strength of the screening shifts the metal-insulator transition to larger values of the bare interaction. The transition remains first order but the coexistence region becomes narrower as the screening effect increases (the two metal-Mott insulator phase boundaries are shown for model (i) at  $\omega_0=5$; for the other cases we show the stability region of the metallic phase).  If the coupling strength is increased into the unphysical overscreening regime, one finds a first order transition to a bipolaronic insulator. 

The right hand panels of Fig.~\ref{phasediagram_i} present the phase diagrams as a function of screened  interaction and coupling strength (parametrized as ratio of unscreened to screened interaction). For $U/U_\text{scr}\sim 2$-$3$,  typical of values found in RPA-type calculations, the critical screened interaction is $\frac{1}{2}$ to $\frac{2}{3}$ of the $U_{c2}$ defined in the unscreened model. The dependence on screening frequency is weak.

\begin{figure}[t]
\begin{center}
\includegraphics[angle=0, width=0.9\columnwidth]{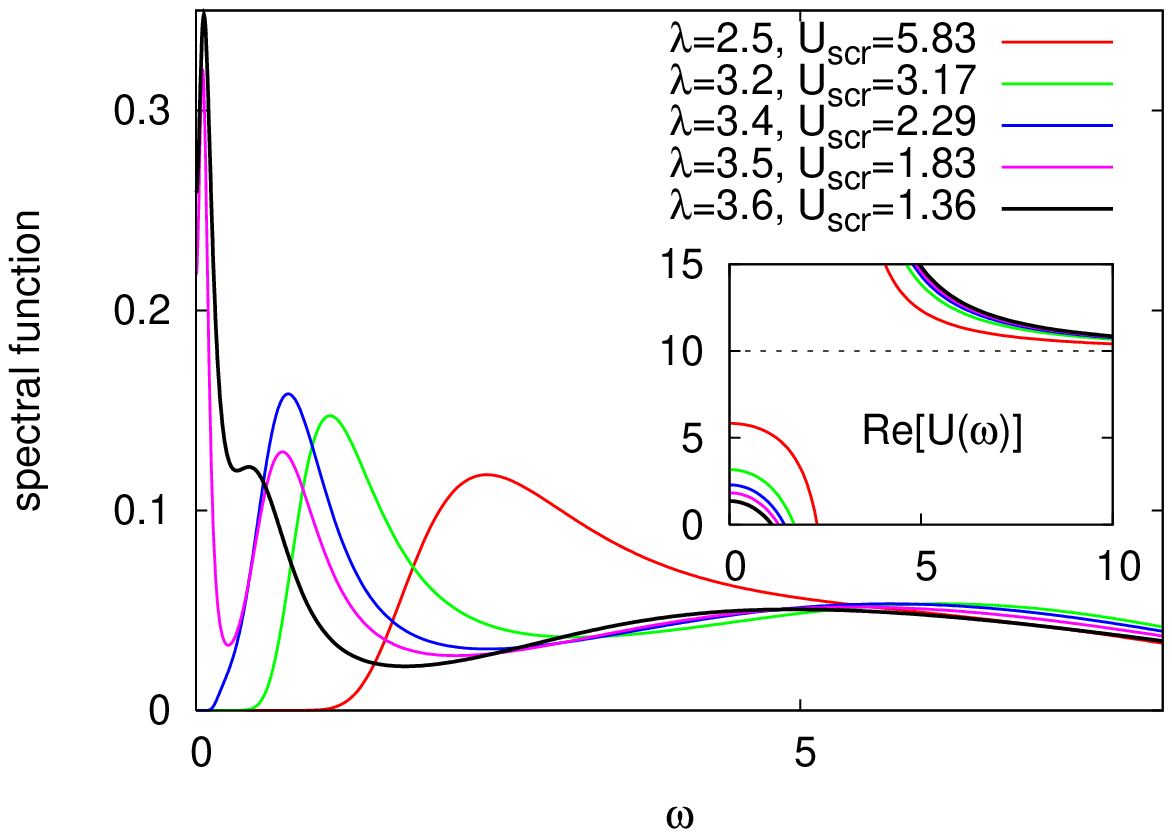}
\includegraphics[angle=0, width=0.9\columnwidth]{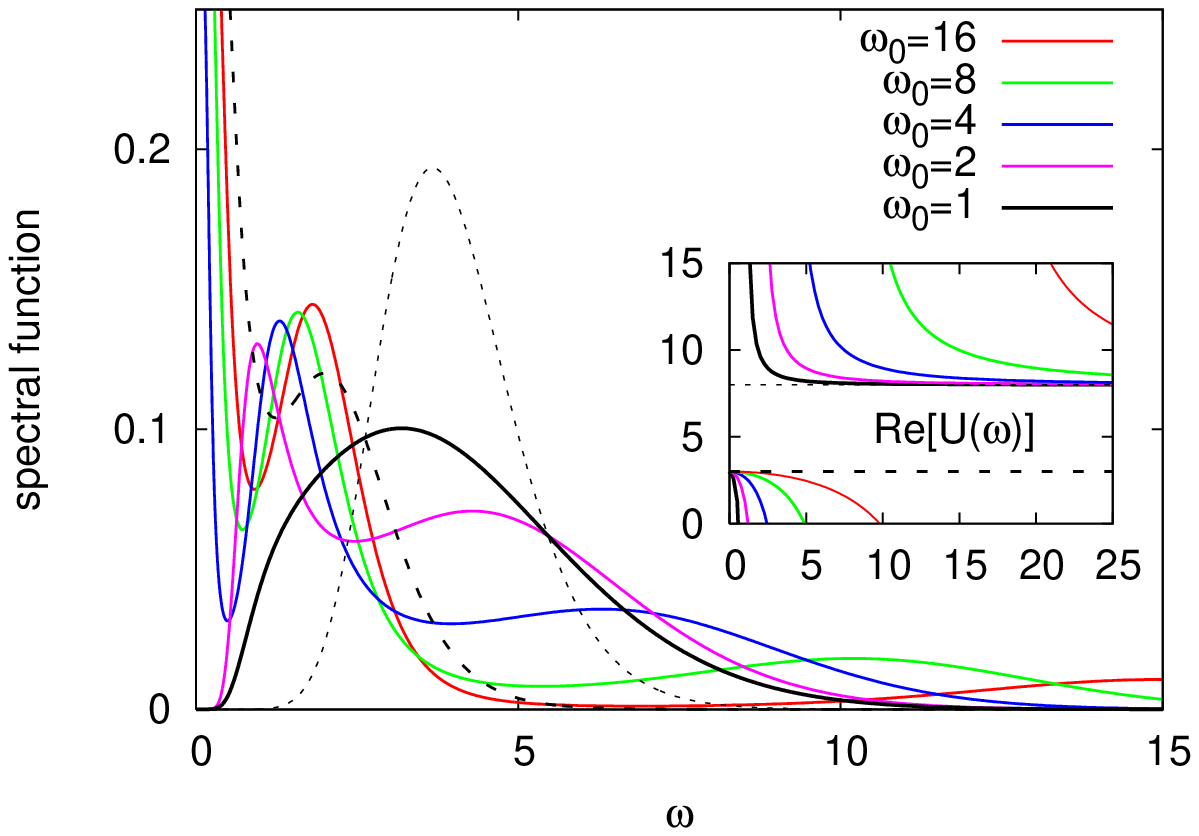}
\caption{Evolution of the spectral function across the metal-insulator transition for model (i). Top panel:  effect of the variation of the screening strength at fixed  $U=10$, $\omega_0=3$. Bottom panel: effect of the variation of the screening frequency at fixed $U=8$ and $U_\text{scr}=3$.  Light (heavy) dashed line: spectrum of the unscreened model for $U=8$ ($U=3$). Inset: frequency dependent interaction $\text{Re} U(\omega) = U+\text{Re} W(\omega)$.
}   
\label{Mott}
\end{center}
\end{figure}

We next consider the effect of screening on the  electron spectral functions. In the model without screening the spectral function is characterized by peaks at $\omega=0$ (if the model is in the metallic phase) and $\omega\simeq\pm U/2$. Figure~\ref{Mott} shows that the situation is quite different in the screened case. The top panel presents the changes that occur as the screening strength is varied at fixed bare interaction $U=10$ and screening frequency $\omega_0=3$. A multipeak structure is evident, with a broad high energy feature at an energy set by a combination of the bare $U$ and $\lambda$, a lower energy sideband  and, in the metallic case, a (split) peak at $\omega=0$. While the peak position of the lower energy sidebands varies roughly in parallel with the screened $U$ its energy is  a parameter-dependent fraction of $U_\text{scr}/2$ (approximately $60\%$).  

The lower panel  of Fig.~\ref{Mott} presents the variation of the spectral function with screening frequency  for  fixed bare and screened interactions $U=8>U_{c2}$, $U_\text{scr}=3<U_{c2}$.  For reference the spectrum in the absence of screening is also shown. At the lowest value of the screening frequency, $\omega_0=1$, the model is in its insulating phase and the spectrum is a broadened version of that of the unscreened insulator.  For all other screening frequencies,  the model is in the metallic phase. We see that the positions of both the lower energy sideband and the  high energy peak increase as the  screening frequency is increased.  At the highest screening frequency the lower energy portion of the spectrum begins to approach that expected in the antiadiabatic limit, with a peak near $U_\text{scr}/2$. We therefore interpret the lower energy feature as the ``screened $U$" sideband, but note that in general its energy does not yield a good measure of $U_\text{scr}$.  Even at the highest screening frequency ($\omega_0=4D=16$),  substantial spectral weight exists at high energies ($\omega\approx\omega_0$).   

\begin{figure}[t]
\begin{center}
\includegraphics[angle=0, width=0.9\columnwidth]{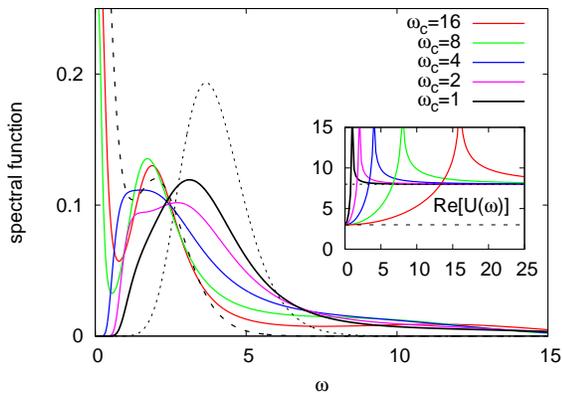}
\caption{
Model (ii): evolution of the spectral function for $U=8$, $U_\text{scr}=3$ and indicated values of $\omega_c$. 
Light (heavy) dashed line: spectrum of the unscreened model for $U=8$ ($U=3$). Inset: $\text{Re} U(\omega) = U+\text{Re} W(\omega)$.
}   
\label{hubbardbands_alpha}
\end{center}
\end{figure}

Figure~\ref{hubbardbands_alpha} shows spectra computed for the Ohmic screening model. The spectra are qualitatively similar to those computed for the plasmon model, but the high energy feature appears as a broad tail and not as a separate peak. 

In summary, we have described a simple and efficient algorithm to treat fermionic lattice models with arbitrary frequency dependent interactions within single-site DMFT.  The only restrictions are that the external screening degrees of freedom may be represented as non-interacting bosons,  have an analytically known commutator with electron creation operators and couple to a quantity which is conserved by the on-site Hamiltonian (this last restriction prevents a direct application of our formalism to clusters). While we have presented results for the  one-band Hubbard model, we emphasize that our method is applicable to multi-orbital models with general Slater-Kamanori interactions. The method therefore opens the door for efficient DMFT simulations of the properties of strongly correlated compounds with arbitrary energy dependence of the interaction parameters, as well as self-consistent GW+DMFT calculations. We also note that the extended-DMFT variants of dynamical mean field theory lead to bosonic problems with a structure very similar to the problem we have considered \cite{Chitra01}. 

The frequency dependent $U(\omega)$ may be characterized by three numbers: a bare (unscreened) interaction, a screened interaction, and a screening frequency. If the screening frequency is very high compared to both the conduction electron bandwidth and the screened interaction, then the physics is well described by an effective model with an instantaneous interaction equal to the screened $U_\text{scr}$. If the screening frequency is very low compared to these scales, then one recovers the familiar electron-phonon physics. We have shown here that the crossover between the two  regimes is very broad. Over wide parameter ranges the dynamical nature of the interaction is important: a model with effective interaction equal to the screened one provides poor estimates of the location of the metal-insulator transition, the value of the Mott Hubbard gaps and the locations and line shapes of the shakeoff features in the spectral function. In particular, the peak positions in the spectal functions do not provide quantitative estimates of either the screened or unscreened $U$ values.   

Our work suggests several directions for future research. Application of the method to real materials is in progress. 
Our approach may also be useful as a solver for E-DMFT calculations. Finally, we note that Assaad and Lang have presented a method for treating bosons within the weak coupling formalism \cite{Assaad07}. While the weak coupling approach is less efficient than the hybridization expansion for single-site (multiorbital) calculations,  it can be applied to cluster dynamical mean field theory. An extension of Assaad's method to investigate the effects of screening in clusters would be of interest.

{\it Acknowledgements} The spectral functions have been obtained using a maximum entropy code implemented by A. Comanac. The calculations were performed on the Brutus cluster at ETH Zurich using a code based on ALPS \cite{ALPS}. PW is supported by SNF-PP002-118866 and AJM acknowledges support from NSF-DMR-0705847.

\end{document}